\documentclass[a4paper,fleqn,usenatbib]{mnras}

\usepackage{newtxtext,newtxmath}
\usepackage[T1]{fontenc}
\usepackage{ae,aecompl}

\usepackage{graphicx}	% Including figure files
\usepackage{amsmath}	% Advanced maths commands
\usepackage{amssymb}	% Extra maths symbols
\usepackage{booktabs}
\usepackage{multirow}
\usepackage{caption}
\usepackage{subfigure}

\title[Emission lines in the atmosphere of WD0137-349B]{Emission lines in the atmosphere of the irradiated brown dwarf WD0137-349B}
\author[E. S. Longstaff et al.]{E. S. Longstaff$^{1}$\thanks{E-mail: el139@le.ac.uk}, S. L. Casewell$^{1}$, G. A. Wynn$^{1}$, P. F. L. Maxted$^{2}$, Ch. Helling$^{3}$ \\
$^{1}$Department of Physics and Astronomy, University of Leicester, University Road, Leicester LE1 7RH, UK \\
$^{2}$Department of Physics and Astrophysics, Keele University, Keele, Staffordshire, ST5 5BG, UK \\
$^{3}$School of Physics and Astronomy, University of St Andrews, St Andrews KY16 9SS, UK
}

\date{Accepted XXX. Received YYY; in original form ZZZ}

\pubyear{2016}

\begin{document}
\label{firstpage}
\pagerange{\pageref{firstpage}--\pageref{lastpage}}
\maketitle

\begin{abstract}

\noindent We present new optical and near-infrared spectra of WD0137-349; a close white dwarf - brown dwarf non-interacting binary system with a period of $\approx$114\,minutes. We have confirmed the presence of H$\alpha$ emission and discovered He, Na, Mg, Si, K, Ca, Ti, and Fe emission lines originating from the brown dwarf atmosphere. This is the first brown dwarf atmosphere to have been observed to exhibit metal emission lines as a direct result of intense irradiation. The equivalent widths of many of these lines show a significant difference between the day and night sides of the brown dwarf. This is likely an indication that efficient heat redistribution may not be happening on this object, in agreement with models of hot Jupiter atmospheres. The H$\alpha$ line strength variation shows a strong phase dependency as does the width. We have simulated the Ca II emission lines using a model that includes the brown dwarf Roche geometry and limb darkening and we estimate the mass ratio of the system to be 0.135$\pm$0.004. We also apply a gas-phase equilibrium code using a prescribed {\sc drift-phoenix} model to examine how the chemical composition of the brown dwarf upper atmosphere would change given an outward temperature increase, and discuss the possibility that this would induce a chromosphere above the brown dwarf atmosphere.
\end{abstract}


\section{Introduction}

Brown dwarfs occupy the mass range between low mass stars and high mass planets; they form like stars through molecular cloud fragmentation but never reach the hydrogen burning mass limit of $\approx$ 0.072M$_\odot$ (\citealt{burrows93, chabrier97}). This lack of hydrogen fusion throughout their evolution results in different atmospheric properties to that of stars. In some respects there is a strong similarity between brown dwarf and planetary atmospheres such as Jupiter (e.g. \citealt{helling14b}). The dividing line on this continuum of objects is the deuterium burning mass limit which occurs above 13\,M$_{\rm Jup}$. This rule was adopted by the Working Group on Extrasolar Planets of the International Astronomical Union in 2002 (\citealt{boss07, spiegel11}). In order to draw clear parallels between planets and brown dwarfs they need to be observed in similar environments i.e. close in to their host star.

Brown dwarf companions that orbit their main sequence star host within 3\,AU are rare compared to planetary or stellar companions to main sequence stars \citep{gret&line06}. This is known as the brown dwarf desert and does not appear to extend out to larger orbital radii \citep{metchev06}. This desert is thought to exist due to the difficulties in forming extreme mass ratio binaries (q $\sim$ 0.02 -- 0.1), although observational biases may play a part too (see Fig. 2 in \citealt{burgasser07}). This is not the case for exoplanets that are thought to form through core accretion in a proto-planetary disc \citep{armitage10}.

There have been a number of brown dwarf companions detected around white dwarf stars but such systems are even rarer than brown dwarfs around main sequence stars with only 0.5\% of white dwarfs having a brown dwarf companion \citep{steele11}. These systems are thought to form through post-common envelope evolution. This is the process by which the white dwarf progenitor evolves off the main sequence, expands, and engulfs the brown dwarf companion. The brown dwarf loses orbital angular momentum to the envelope thus causing the brown dwarf to spiral in towards the core and the envelope to be ejected \citep{politano04}. This results in a detached system with a white dwarf and a close, tidally locked brown dwarf companion known as a post common envelope binary (PCEB).

Recent infrared studies have provided candidate systems (e.g. \citealt{debes11, girven11, steele11}) that have been identified by looking for infrared excesses in SDSS spectra. Only a handful of the candidate systems have been confirmed as PCEBs: GD1400 (WD+L6, P=9.98\,hrs; \citealt{farihi04, dobbie05, burleigh11}), WD0137-349 (WD+L6-L8, P=116\,min; \citealt{maxted06, burleigh06, casewell15}), WD0837+185 (WD+T8, P=4.2hrs; \citealt{casewell12}), NLTT5306 (WD+L4-L7, P=101.88\,min; \citealt{steele13}), SDSS J141126.20+200911.1 (WD+T5, P=121.73\,min; \citealt{beuermann13, littlefair14}), SDSS J155720.77+091624.6 (WD+L3-5, P=2.27\,hrs; \citealt{farihi17}), SDSS J120515.80-024222.6 (WD+L0, P=71.2\,mins) and SDSSJ123127.14+004132.9 (WD+BD, P=72.5\,mins; \citealt{parsons17}). These systems are thought to be progenitors of cataclysmic variables with a brown dwarf donor (e.g. \citealt{littlefair03, hern16}).

There are advantages to studying irradiated brown dwarfs over hot Jupiters. Firstly the atmospheres of brown dwarfs have been very well characterised (e.g. \citealt{cushrayvac05}) and the models associated with brown dwarf atmospheres are thus more advanced than exoplanet models (e.g. \citealt{helling13, helling14b, marley15}). Secondly brown dwarfs are typically brighter than exoplanets and can be directly observed at wavelengths longwards of 1.2\,$\mu$m. Thirdly, observing brown dwarf or planetary atmospheres around main sequence stars is difficult due to the brightness of their host. A solution to this is to observe these systems in their more evolved form. A brown dwarf emits primarily in the near-infrared whereas the white dwarf contribution dominates at short wavelengths. Due to this there is a high contrast between the brown dwarf and white dwarf making it easier to separate their spectral components. Thus the observation driven study of short period irradiated brown dwarf atmospheres can give us insights into the chemistry of the atmospheres of hot Jupiters.

WD0137-349 has been the subject of several studies over the last decade e.g. \citep{maxted06, burleigh06, casewell15}. The white dwarf was first discovered using low resolution spectroscopy and added to the McCook \& Sion catalog \citep{McCookSion99}. Follow up high resolution spectroscopy by \citet{maxted06} revealed the presence of a brown dwarf companion and concluded that it was not affected by the common envelope phase. 2MASS photometry and observations using the Gemini telescope and the Gemini Near-Infrared Spectrograph revealed an infrared excess. This allowed \citet{burleigh06} to better characterise the brown dwarf and estimate its spectral type to be L6--L8, although these observations were taken of the unirradiated hemisphere of the brown dwarf. The brown dwarf intercepts 1 per cent of the primarily UV white dwarf radiation and is likely to be tidally locked to its 16\,500\,K host. This results in one side being continually irradiated giving the brown dwarf a temperature difference of 500\,K  between its ''night-" and ''day -- side" \citep{casewell15}. \\

Our observations and data analysis are summarised in the next section. In section 3 we present our measured radial velocities and equivalent widths of all the emission lines detected, we also present the refined ephemeris of this system. In section 4 we put WD0137-349B into the context of similar binaries. In section 5 we discuss how we used a {\sc drift-phoenix} model to do a first test of how the chemical composition of the brown dwarf upper atmosphere would change given an outward temperature increase. We discuss the possibility that this would induce a chromosphere above the brown dwarf atmosphere.

\section{Observations and data reduction}

We obtained 78 spectra using the XSHOOTER instrument \citep{xshooterref} mounted at the ESO VLT-UT3 ('Melipal') telescope in Paranal, Chile. We observed in the wavelength range 300.0 -- 2480.0\,nm using the three independent arms: VIS, UVB, NIR. The observations took place on the nights of 2014-August-28 between 04:45 -- 09:45 and 2014-August-29 between 04:35 -- 07:12 universal time. The 78 exposures, each 300\,s long, were taken covering 2.4 orbits on the first night and 1.3 orbits on the second night. This gives us 21 spectra per full orbit. The weather conditions on the first night were good with an airmass ranging between 1.02 -- 1.33 and seeing of 0.98\,arcsec. The second night the weather was not as good due to a high wind of 13\,ms$^{-1}$ at the start of the night and slowing to 8\,ms$^{-1}$. The airmass ranged between 1.02 -- 1.58 and the seeing between 0.63 -- 1.1\,arcsec. The data and standards were taken in nodding mode and were reduced using \textsc{gasgano} (v2.4.3; \citealt{gasgano}) following ESO's XSHOOTER pipeline (v2.5.2).

We see H$\alpha$ absorption and emission features that move in anti-phase over the orbital period of the system. The data were normalised and phase-binned using Tom Marsh's \textsc{molly} software. The H$\alpha$ absorption feature has a narrow core and broad wings. To fit this line we followed the method from \cite{casewell15} and used three Gaussian profiles: two for the wings and one for the core. The best fitting widths were 0.149, 2.360, and 8.30\,nm and the depth of the narrowest profile was fixed. A fourth Gaussian was used to model the H$\alpha$ emission, height and width were free to roam. This allowed us to measure the radial velocities of the brown dwarf and white dwarf independently. For the other emission lines there are no features coming from the white dwarf at those wavelengths therefore only one Gaussian was necessary to model these emission lines. 

\section{Results}

\begin{table*}		%Parameter results - THE BIG TABLE
\begin{center}
\begin{tabular}{l r r r r r}
\toprule
	 										& Wavelength				& Semi-amplitude 	& Systemic velocity			& Day-side EW 					& Night-side EW\\
											& $\lambda$ (\AA)		& K (kms$^{-1}$)	&$\gamma$ (kms$^{-1}$)	&(\AA) 								&(\AA)\\
\midrule
\multirow{2}{*}{H$\alpha$}	& 6562.76 (ab)		&$28.3\pm0.4$		&$18.4\pm0.3$					&	-										& -  \\
												& 6562.76 (em)		&$-192.4\pm1.2$	&$4.5\pm0.9$					& $-0.79\pm0.021$ 			&$-0.03\pm0.021$  \\
\midrule
\multirow{2}{*}{H$\beta$}	& 4861.30 (ab)		&$28.9\pm0.9$		&$6.8\pm0.3$					& -										& - \\
												& 4861.30 (em)		&$-189.5\pm5.2$	&$2.1\pm3.7$					&$-0.29\pm0.030$ 			&$-0.09\pm0.030$ \\
\midrule
H I										& 10938.17			&$-184.8\pm8.7$		&$-14.8\pm6.2$				& -										& - \\
\midrule
He I										& 21655.34			&$-119.3\pm50.3$	&$1.7\pm24.2$					& $-0.66\pm0.1$ 				& $0.24\pm0.1$\\
\midrule
\multirow{4}{*}{Na I} 		& 5889.95				&$-188.2\pm2.4$		&$2.9\pm1.7$	 				&$-0.20\pm0.021$			&$-0.02\pm0.021$\\
											& 5895.92				&$-185.7\pm2.5$		&$6.6\pm1.8$	 				&$-0.16\pm0.021	$			&$0.04\pm0.021$\\
											& 8183.27				&$-191.2\pm7.0$		&$1.0\pm4.8$					&$-0.13\pm0.021$			&$0.03\pm0.021$\\ 					
											& 8194.81				&$-184.8\pm6.1$		&$-4.4\pm4.2$					&$-0.19\pm0.021	$			&$0.06\pm0.021$\\
\midrule
\multirow{4}{*}{Mg I} 		& 5167.32				&$-188.1\pm2.4$		&$-3.5\pm1.7$ 					&$-0.15\pm0.030	$			&$-0.05\pm0.030$\\
						 					& 5172.68				&$-186.7\pm2.3$		&$-3.8\pm1.6$ 					&$-0.17\pm0.030	$			&$-0.09\pm0.030$\\
	 										& 5183.60				&$-187.9\pm1.7$		&$-6.4\pm1.2$ 					&$-0.14\pm0.030	$			&$-0.07\pm0.030$\\
											& 8806.76				&$-191.4\pm5.0$		&$-23.0\pm3.5$				&$-0.40\pm0.021	$			&$-0.03\pm0.021$\\							
\midrule
\multirow{3}{*}{Si I}			& 10585.14			&$-171.8\pm12.7$	&$10.9\pm9.0$ 				&$-0.32\pm0.1$ 				& $0.04\pm0.1$\\	
											& 10868.79			&$-182.5\pm5.0$		&$0.5\pm3.7$ 					&$-0.09\pm0.1$ 				& $0.23\pm0.1$\\
											&	12031.51 			&$-187.1\pm4.8$		&$-13.2\pm3.7$ 				&$ -0.29\pm0.1$				& $0.03\pm0.1$\\
\midrule				 		
K I				  						& 7698.96				&$-187.5\pm2.0$		&$7.1\pm1.4$					&$-0.14\pm0.021	$			&$0.00\pm0.021$\\
\midrule
Ca I 									& 12816.04 			&$-184.8\pm6.3$		&$19.8\pm4.2$ 				& - 										& - \\
\midrule
\multirow{4}{*}{Ca II} 		& 3933.66				&$-185.0\pm4.2$		&$-1.2\pm2.9$ 					&$-0.08\pm0.030	$			&$0.05\pm0.030$\\
											& 8498.02				&$-186.5\pm0.8$		&$6.3\pm0.6$					&$-0.46\pm0.021	$			&$-0.01\pm0.021$\\
 											& 8542.09				&$-185.6\pm1.1$		&$7.1\pm0.8$					&$-0.75\pm0.021	$			&$0.00\pm0.021$\\
											& 8662.14				&$-186.9\pm1.0$		&$5.1\pm0.7$					&$-0.73\pm0.021	$			&$-0.05\pm0.021$\\
\midrule
\multirow{3}{*}{Ti I} 			& 11752.31 			&$-177.6\pm4.4$		&$13.1\pm3.0$					&$-0.05\pm0.1$ 				&$-0.06\pm0.1$ \\
											& 11982.96 			&$-191.0\pm3.5$		&$8.2\pm2.7$ 					&$-0.14\pm0.1$ 				&$-0.05\pm0.1$\\
											& 19860.98			&$-158.7\pm8.5$		&$-6.2\pm4.6$ 					&$-0.44\pm0.1$ 				&$0.05\pm0.1$\\
\midrule
\multirow{5}{*}{Fe I}			& 8327.05				&$-186.3\pm2.2$		&$3.6\pm1.6$					&$-0.11\pm0.021$				&$0.003\pm0.021$ \\	
											& 8334.53				&$-185.7\pm4.6$		&$19.7\pm3.2$					&$-0.10\pm0.021$				&$0.04\pm0.021$ \\
											& 8387.77				&$-185.2\pm2.2$		&$7.7\pm1.6 $					&$-0.16\pm0.021$				&$-0.01\pm0.021$ \\
											& 8688.62				&$-188.9\pm1.4$		&$7.2\pm1.0 $					&$-0.15\pm0.021$				&$0.01\pm0.021$ \\
											& 8824.22				&$-186.4\pm2.5$		&$6.6\pm1.7 $					&$-0.05\pm0.021$				&$-0.01\pm0.021$ \\	
											& 10682.39			&$-193.9\pm16.8$	&$40.3\pm12.1$				&$-0.10\pm0.1$ 					&$-0.05\pm0.1$ \\
											& 11970.50			&$-186.5\pm11.6$	&$2.1\pm7.4$ 					&$-0.04\pm0.1$ 					&$-0.10\pm0.1$\\
											& 13120.47			&$-160.6\pm27.5$	&$-31.3\pm17.1$ 				&$-0.07\pm0.1$ 					&$-0.18\pm0.1$\\
											& 13147.92			&$-210.1\pm10.6$	&$5.0\pm6.5$					&$-0.07\pm0.1$ 					&$0.07\pm0.1$\\
											& 14876.54			&$-219.6\pm25.4$	&$14.0\pm15.5$ 				&$-0.77\pm0.1$ 					&$-0.1\pm0.1$\\
											& 15022.67			&$-185.0\pm5.3$		&$6.7\pm3.7$ 					&$-0.13\pm0.1$ 					&$0.02\pm0.1$\\
											& 15036.47			&$-168.7\pm18.0$	&$3.9\pm12.4$ 				&$-0.14\pm0.1$ 					&$-0.24\pm0.1$\\
											& 19771.29			&$-180.5\pm15.5$	&$9.0\pm7.9$					&-  											& - \\
\midrule								
Fe II									& 8434.49				&$-198.2\pm3.9$		&$17.8\pm2.8$					&$-0.15\pm0.021	$				&$-0.07\pm0.021$\\
											& 10690.51			&$-180.8\pm4.7$		&$-0.8\pm3.5$					&$-0.15\pm0.1$ 					&$0.02\pm0.1$ \\
											& 11825.65			&$-182.2\pm10.8$	&$18.0\pm7.2$					&$-0.03\pm0.1$ 					&$0.05\pm0.1$ \\
											& 15885.67			&$-171.0\pm15.4$	&$2.1\pm10.1$ 				& - 											& - \\
											& 17106.54			&$-175.0\pm12.7$	&$10.6\pm7.8$					&$-0.17\pm0.1$ 					&$-0.20\pm0.1$\\

\bottomrule
\end{tabular}
\end{center}
\caption{The best fit radial velocity parameters (cf. equation~\ref{eq:rv}). The period and zero ephemeris were determined by fitting the white dwarf radial velocity data and can be found in Table~\ref{table:masses}. $P$ and $T_0$ and were then held constant for all subsequent emission lines. The equivalent widths of the emission lines were determined, where possible, from the line profiles.}
\label{table:orbprop}
\end{table*}

\begin{figure}
\begin{center}
\includegraphics[height=8.0cm, angle=270]{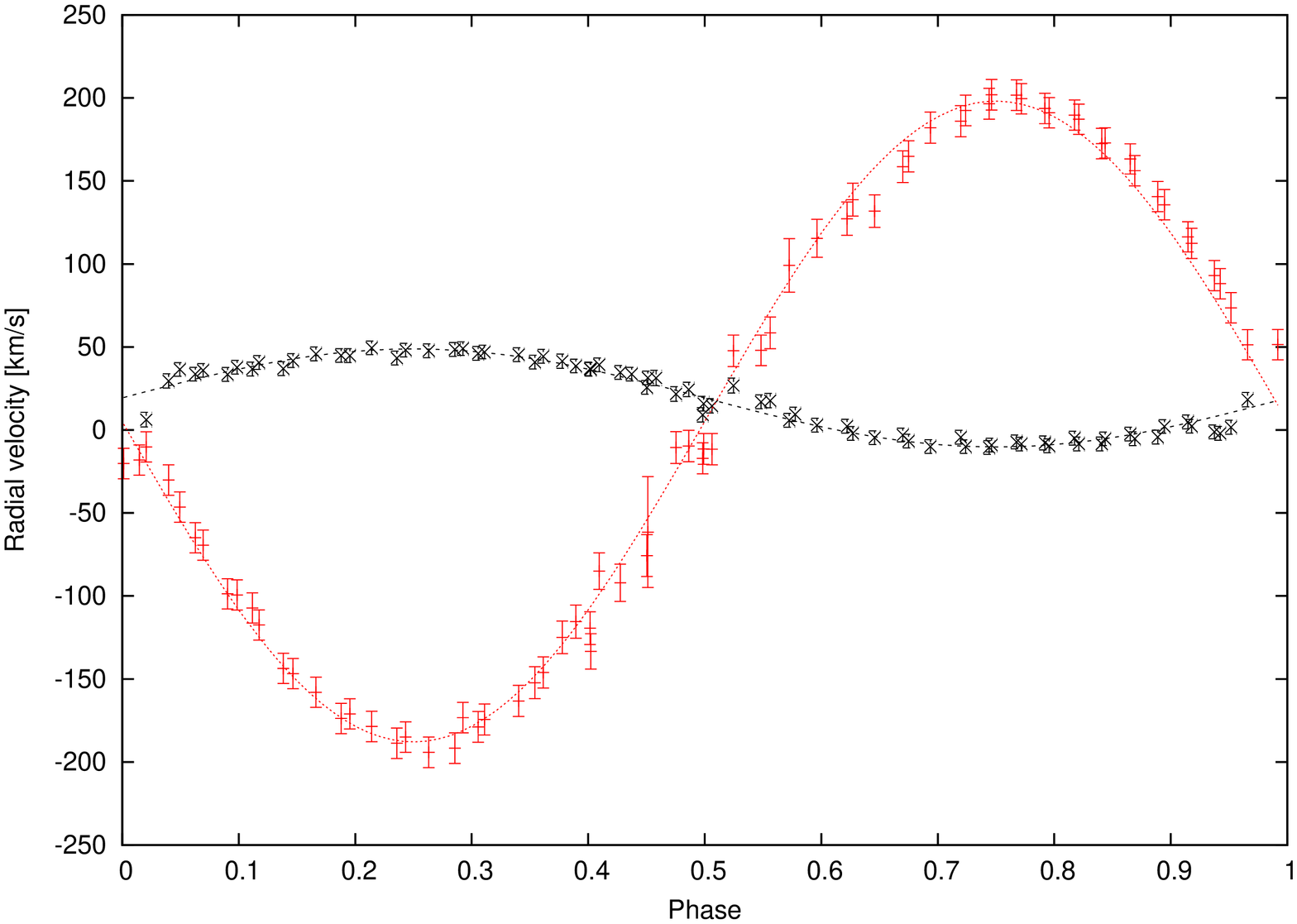}
\end{center}
\caption{The radial velocity values of the H$\alpha$ emission (red '+') and absorption (black 'x'). These have been fit with using equation~(\ref{eq:rv}) the values from table~\ref{table:orbprop}.}
\label{fig:bd_plot}
\end{figure}

\noindent The data from the two nights were analysed together and separately and no significant difference was found between the two. Therefore we will use the analysis of both nights together. \\

\subsection{Refining the ephemeris}

The radial velocities of the white and brown dwarf were extracted from our multi-Gaussian fits. These were then fitted with a sine function of the form,

\begin{equation}
V_r = \gamma + K\sin{2\pi\phi},
\label{eq:rv}
\end{equation}

\noindent where $V_r$ is the measured radial velocity in kms$^{-1}$, $\gamma$ is the systemic velocity of the object in kms$^{-1}$ and $K$ is the semi-amplitude in kms$^{-1}$. Phase, $\phi$, is calculated by subtracting the zero-point of the ephemeris, $T_0$, from the HJD, $t$, and dividing by the period, $P$,

\begin{equation}
\phi = \frac{t - T_0}{P}.
\label{eq:phase}
\end{equation}

\noindent To determine $\gamma$, $K$, $P$, and $T_0$ we used a least squares method to fit equations~\ref{eq:rv} and~\ref{eq:phase} to the H$\alpha$ absorption and emission lines in our 78 spectra. Any points more than 3$\sigma$ away from the model were excluded to improve the fit and 1\,kms$^{-1}$ was added in quadrature to the standard statistical errors until a reduced $\chi^2$ of 1 was achieved. The error bars shown on Figure~\ref{fig:bd_plot} reflect this and the solid lines are the best fit from the least--squares fit. The period and zero ephemeris were then held constant when fitting the other emission lines and their semi-amplitudes and systemic velocities are listed in Table~\ref{table:orbprop}.

\subsection{Spectral features}

\begin{figure*}
\subfigure[The H$\alpha$ emission (light grey) and absorption (dark grey) line.]{\includegraphics[height=8.0cm, width=8.0cm]{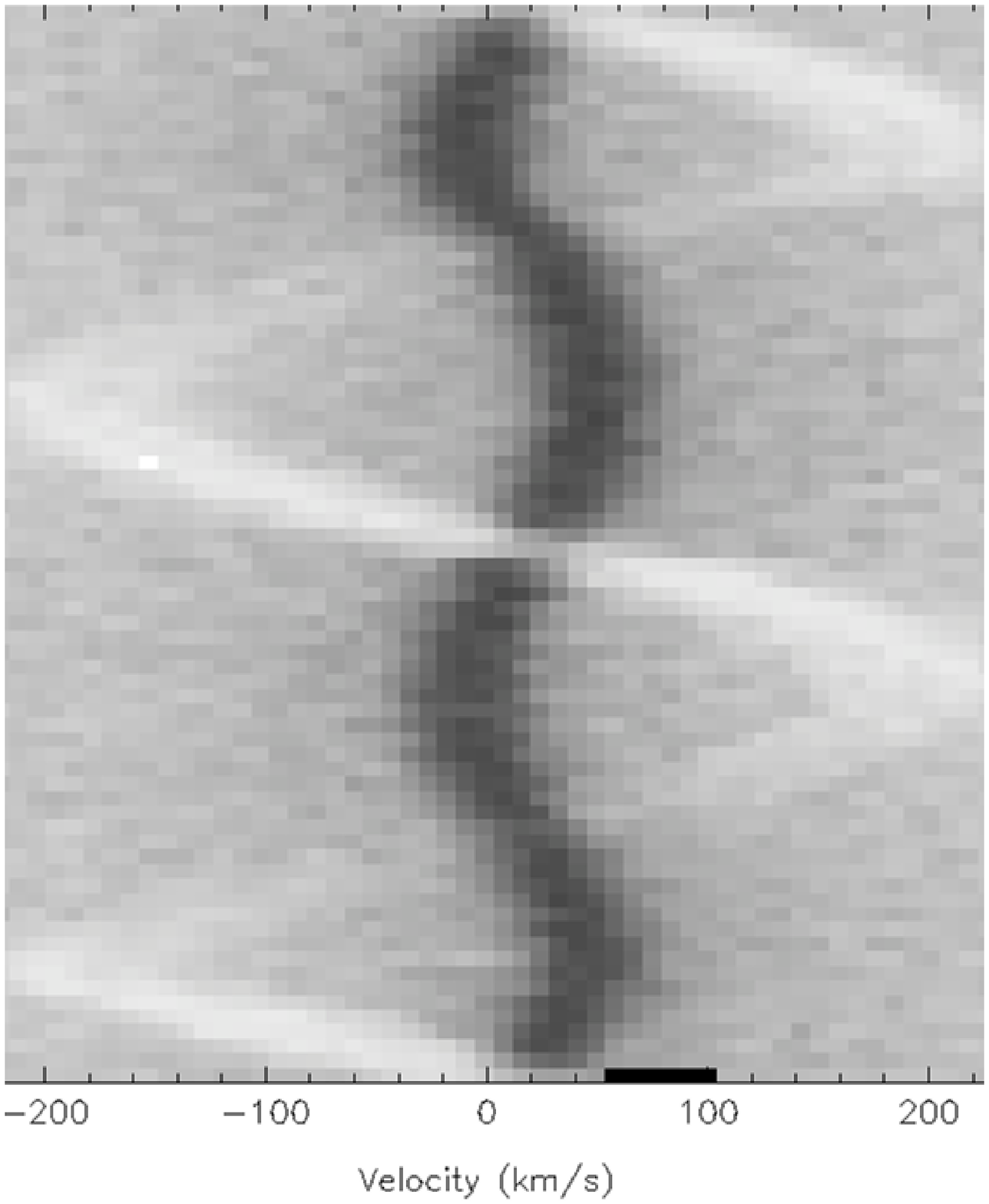}}
\subfigure[The 8662.14\,\AA\, Ca II emission line (light grey).]{\includegraphics[height=8.0cm, width=8.0cm]{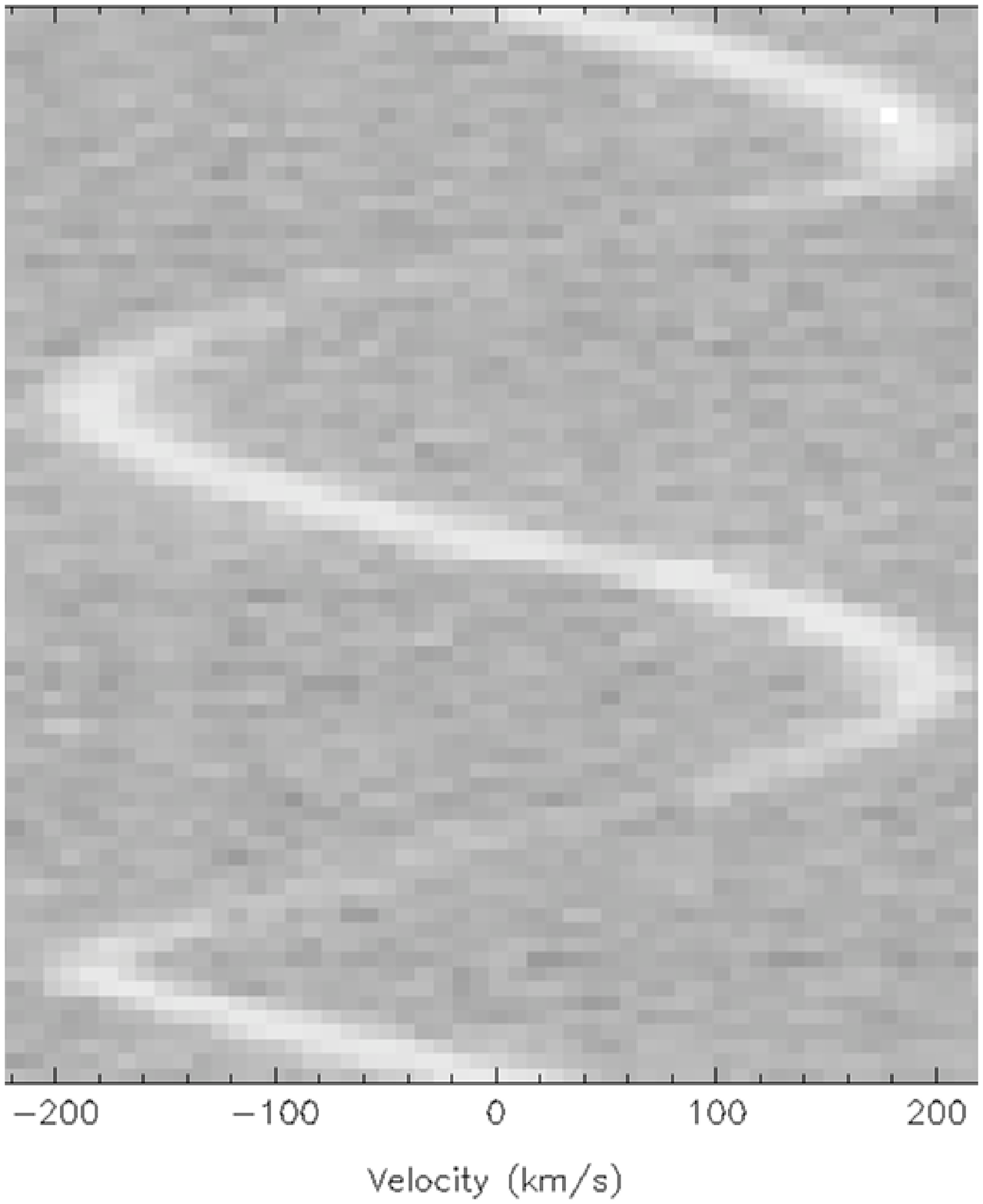}}
\caption{Trailed spectra demonstrating the binary nature of the system. The H$\alpha$ emission line is in anti-phase with the absorption feature. The Ca II emission closely resembles the H$\alpha$ emission indicating they originate from the same place: the brown dwarf.}
\label{fig:bn_trail}
\end{figure*}

We have identified several other emission lines: He~I, Na~I, Mg~I, Si~I, K~I, Ca~I \& II, Ti~I, and Fe~I \& II (see Table~\ref{table:orbprop}). The results for each line are very similar, so for simplicity we will analyse H$\alpha$ and use Ca II (866.214\,nm) for comparison.

Figure~\ref{fig:bn_trail} shows the trailed spectra of H$\alpha$ demonstrating the binary nature of the system. The adjacent trailed spectra of the Ca~II emission line highlights the similarity between the H$\alpha$ and Ca~II emission features. These emission features disappear briefly when the ''night"-side of the brown dwarf is facing the observer ($\phi=0.5$)\footnote{For consistency we follow previous studies of WD0137-349 in adopting the unconventional phasing of this system and define phase 0 as when the brown dwarf is furthest from the observer.}. Therefore it is likely the emission is only coming from the heated side of the brown dwarf. This is happening in both trails however it is slightly less obvious in the H$\alpha$ trail due to the absorption feature.

In addition to our search for emission features, we also looked for absorption lines which may be present in the system either from the brown dwarf -- which has been seen in sdB systems \citep{wood99} -- or intrinsic to the white dwarf, indicating it is a DAZ. The white dwarf in this system is within the temperature range of 16\,000 -- 20\,000\,K suggested by \citet{kilic06} to be cool enough to potentially show metal absorption lines in its spectrum, however it does not. Only the Balmer lines are present coming from the white dwarf. There is also no emission in phase with the white dwarf therefore no mass transfer or gas circulation can be inferred.

\section{Comparison to similar systems}

%\textbf{and within this small group of binaries WD0137-349 stands out because of the range of emitting species present in the companion atmosphere.}

WD0137-349B is unique because it is the first irradiated brown dwarf to exhibit He, Na, Mg, Si, K, Ca, Ti, and Fe emission lines in its atmosphere. The radial velocities of these lines confirm that they originate from the brown dwarf and are most likely a direct result of irradiation from the 16\,500\,K white dwarf host. Below we contrast the observable properties of WD0137-349B to other, similarly irradiated brown dwarf and low-mass stellar companions to white dwarfs in short period binaries.

A large number of PCEBs have low mass stellar companions and Balmer line emission has been detected in a number these.  For example, LTT 560 displays H$\alpha$ emission with dual components, one of which is due to stellar activity on the M dwarf \citep{tappert07}. Ca I \& II, Na I, K I, Fe I and Ti were observed in the secondary's atmosphere as absorption \citep{tappert11}. Irradiation induced emission lines outside of the Balmer series are less common but have been seen. For example Ca II and Na I emission in GD 448 \citep{maxted98} and He I and Mg II emission in NN Serpentis \citep{parsons10}.

A handful of PCEBs have substellar companions but emission from these is seen less frequently. H$\alpha$ emission has been detected in WD0137-349 before \citep{maxted06}; it has also been seen in two other systems: SDSS J120515.80-024222.6 and SDSS J155720.77+091624.6. The former is an eclipsing system with a short period (71.2\,mins) and a $\sim$23,500\,K white dwarf \citep{parsons17} and the latter has a longer period (2.27 hours) and a 21,800\,K white dwarf \citep{farihi17}.

Whilst WD0137-349B is the first brown dwarf companion to exhibit irradiation induced metal emission lines, irradiation has been observed to induce temperature differences across the surfaces of brown dwarf companions. For example SDSS J141126.20 +200911.1 displays a flux excess in the K$_s$ band indicative of a temperature increase on the irradiated side of the brown dwarf \citep{littlefair14} and the cataclysmic variable SDSS J143317.78 +101123.3 shows a $\sim$ 200\,K temperature difference across the surface of the brown dwarf \citep{hern16}. WD0137-349 has a temperature difference of 500\,K between the day-- and night--side of the brown dwarf \citep{casewell15}.

These irradiated white dwarf -- brown dwarf systems do not all manifest the same observable properties. In order to understand this we need to look at the atmospheres of isolated brown dwarfs to establish a base from which we can interpret irradiated brown dwarfs.

%One would expect that shorter orbits with hotter white dwarfs, thus more intense irradiation, would lead to stronger Balmer line emission and metal emission like we see in WD0137-349.

H$\alpha$ emission as been seen in the atmospheres of "hyperactive" brown dwarfs. \citet{pineda16} characterised the H$\alpha$ emission in three such objects objects: 2MASS J00361617+1821104 (2MASS 0036), 2MASS J17502484-0016151 (2MASS 1750), and SDSS J042348.57-041403.5 (SDSS 0423). 2MASS 0036 is an L dwarf with variable H$\alpha$ which is thought to be linked to magnetic interaction with a companion. The H$\alpha$ emission isn't as strong in 2MASS 1750 which has been classified as a L5.5 brown dwarf and SDSS 0423 is an L6+T2 brown dwarf binary with the H$\alpha$ emission coming from the L dwarf. The detection of H$\alpha$ emission is linked to chromospheric activity in brown dwarf atmospheres. The M9.5 dwarf PC 0025+0447 (M$<$0.06M$_\odot$) discovered by \citet{schneider91} displays consistent H$\alpha$ emission. This emission and the emission of He~I and O~I is thought to be due to coronal activity on the brown dwarf \citep{martin99}. 2MASS J04183483+2131275 a L5$\pm$0.5 brown dwarf is the first, in the Hyades open star cluster, to display chromospheric H$\alpha$ emission \citep{perez-garrido17}. This activity is unusual and is not expected from a brown dwarf of this age. A double degenerate binary system, 2MASS J13153094-2649513 (L5 + T7; \citealt{burgasser11}), shows H$\alpha$, Na~I, and K~I emission coming from the L dwarf \citep{fuhrmeister05}. The lack of irradiation suggests the emission is non-thermal, the cause is thought to be the presence of a chromosphere driven by an unusually strong magnetic field \citet{burgasser11}. 

WD0137-349B displays metal emission lines akin to irradiated low-mass stellar companions but is the first sub-stellar companion to do so. So-called hyperactive brown dwarfs have shown similar emission which has been taken as evidence for the presence of a chromosphere. The difference being hyperactive brown dwarf  chromospheres are thought to be generated by unusually strong magnetic fields; whereas WD0137-349B is due to the intense irradiation from the white dwarf. We will discuss the presence of brown dwarf chromospheres with respect to WD0137-349B in the next section.

\section{Discussion}

There are some distinct atmospheric differences between WD0137-349 and field brown dwarfs. For example, molecules such as CrH and FeH are expected to be present in a L type dwarf but we do not detect either of these molecules in our data. Nor do we see any sign of Cr, Rb, or Cs which are also expected in a substellar atmosphere. These elements are raised in the atmosphere through the process of convection \citep{oppenheimer98}, so their non-detection suggests that irradiation is hindering the convective processes that would normally occur in a brown dwarf atmosphere.

The metal emission lines we have detected allow us to draw some parallels, for example, Na~I and K~I are expected to be in the atmosphere of a late L-type brown dwarf (\citealt{kirkpatrick99, helling14b}). A brown dwarf of spectral type L7 or later would not be expected to display TiO in its spectrum \citep{kirkpatrick99} and indeed we do not detect any molecules at all. We have however detected Ti I suggesting irradiation may have split this molecule leaving Ti in the upper-atmosphere.

\begin{figure*}
\begin{center}
\vspace{1cm}
\includegraphics[height=10.0cm, width=12.0cm]{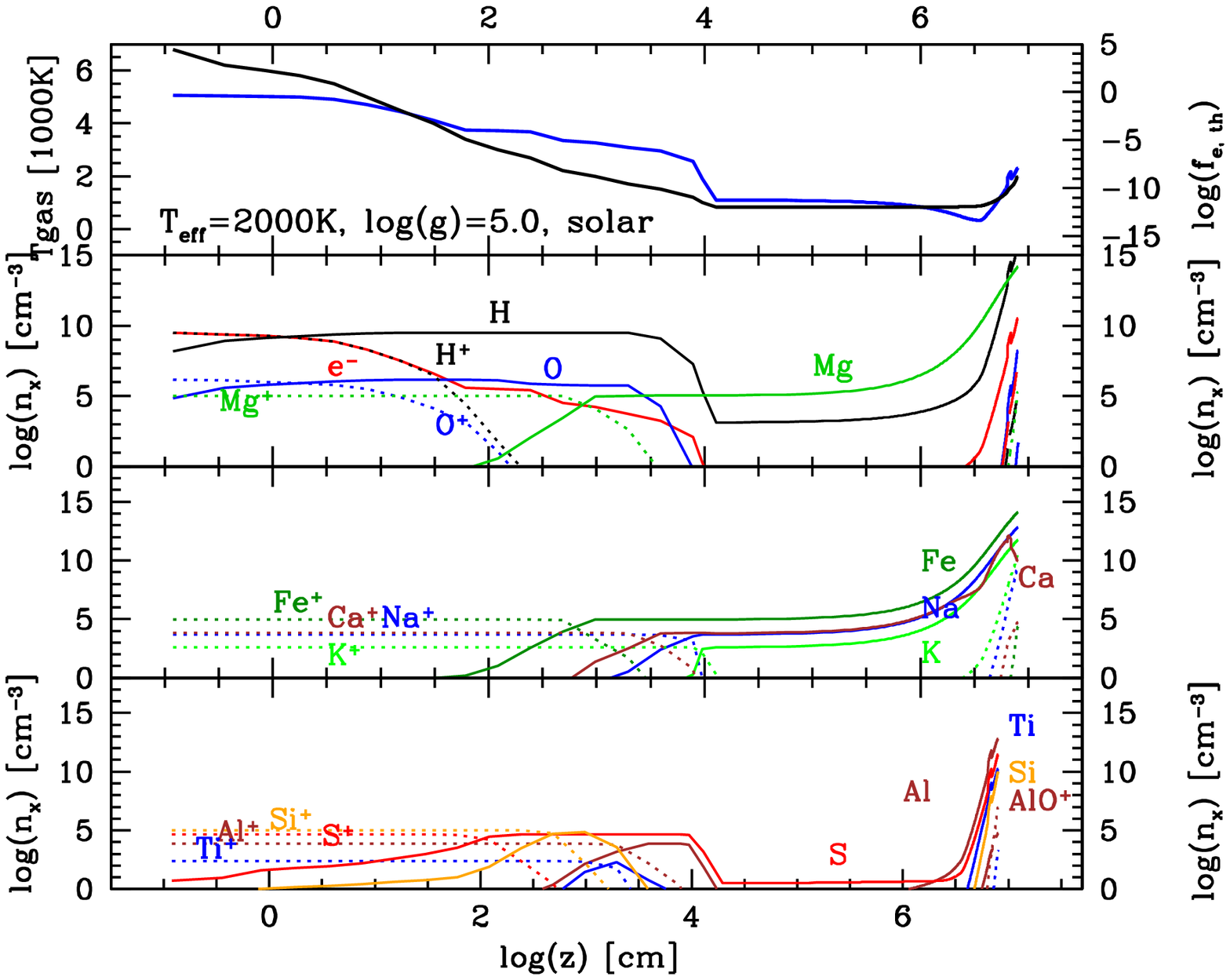}
\vspace{1cm}
\end{center}
\caption{Chemical equilibrium results for a Drift-Phoenix model atmosphere (T$_{\rm eff}$ = 2000\,K, log(g) = 5.0, solar metalicity) with an artificially added chromosphere-like temperature increase for atmospheric heights z\,$<10^4$\,cm (z decreases outwards). The top panel shows the temperature of the brown dwarf (black, steeper line) due to a chromosphere or external energy input and the degree of thermal ionisation (blue, flatter line) that these higher temperatures induce. The second, third, and fourth panels show the chemical equilibrium results plotted for a selected number of atoms (solid lines) and ions (dashed lines) each with a corresponding label. As expected, the inner part of the atmosphere is dominated by the atomic form of an element but an outward increasing temperature (e.g. resulting from a chromosphere or corona) will cause different ions to appear height-dependently.}
\label{fig:helling_plot}
\end{figure*}

Atmospheres of single or binary brown dwarfs are understood to have very low effective temperatures. The gas-phase is therefore dominated mostly by  H$_2$, as well as H$_2$O, CO or CH$_4$, depending on the effective temperature. If the brown dwarf is irradiated by an external source, the atmospheric temperature will rise similar to the outer atmospheres of irradiated planets (see Fig 1 in \citealt{barman04}). \citet{schmidt15} present observations from the BOSS survey that suggest that brown dwarfs have chromospheres and this is corroborated by \citet{sorahana14} who reach a similar conclusion. \citet{rodriguez15} demonstrate that the upper parts of a brown dwarf atmosphere can be magnetically coupled (see their Sect.6.1) despite the low atmospheric temperatures that allow the formation of clouds (e.g. \citealt{helling14b}). Brown dwarfs are known to be convectively active such that magnetic waves could be excited thus causing a heating of the upper, low-density atmosphere; this is similar to what occurs within the Sun (e.g. \citet{mullan16}). \citet{schmidt15} and \citet{sorahana14} both invoke a simplified representation of the chromosphere temperature structure. A chromosphere in WD0137-349B could for example originate from MHD processes like Alfv\'en wave heating but also be a result of the high-energy irradiation from the white dwarf. We have therefore followed a similar approach to these papers to mimic the presence of a chromosphere. This allows us a simple first test of the expected chemical composition in the hot and diluted upper part of the brown dwarf atmosphere.

Figure~\ref{fig:helling_plot} shows a first test of how ions would dominate the chemical composition in a hot upper atmosphere of WD0137-349B. We apply a gas-phase equilibrium code (see \citealt{bilger13}) and use a prescribed {\sc drift-phoenix} model (T$_{\rm gas}$, gas)-structure for T$_{\rm eff}$=2000\,K, log(g)=5.0, [M/H]=solar as input for the inner atmospheric part. We manipulate the temperature in the upper atmosphere to be indicative of a temperature increase caused by a chromosphere or external energy input (black line, top panel, Fig. ~\ref{fig:helling_plot}). The gas phase temperature is now high enough to increase the local degree of thermal ionisation (blue line, top panel) substantially as is suggested by the observations presented here. Panels 2 -- 4 in Figure~\ref{fig:helling_plot} demonstrate which ions would emerge if WD0137-349B exhibited such a rise in local temperature in the upper atmosphere. Various ions dominate over their atomic counterpart at different gas temperatures i.e. at different atmospheric heights. For example, H$^+$ (H~II) is the dominating H-binding species rather high in our example atmosphere where T$_{\rm gas} \approx$ 6000\,K while Fe~II (Fe$^+$) dominates over Fe~I (Fe) at T$_{\rm gas} \approx$ 2100\,K for the given density structure; the uppermost atmospheric layer is now fully ionised with n(H$^+$) = n(e$^-$). While the calculation presented in Fig.~\ref{fig:helling_plot} does only consider collisional ionisation in LTE, it gives an indication of how the upper atmosphere of a brown dwarf would change if irradiation, or another mechanism for forming a chromosphere, caused an outward temperature increase which is necessary to explain the emission features observed for WD0137-349B.

Our simple approach suggests that atomic emission lines should be expected to occur in brown dwarf atmospheres. Irradiated brown dwarfs will, however, have strong temperature gradients in the observable atmosphere similar to irradiated giant gas planets e.g. HD\,189733b \citep{lee16}. Such 'surface' temperature gradients are caused by hydrodynamic transport of atmospheric gas and will cause the local chemistry to change substantially \citep{helling16}. Therefore, a far hotter day-side of an irradiated brown dwarf would show higher ionised elements in emission than the terminator regions, for example. Our data supports this as the strength of the emission lines significantly weaken when the night--side of the brown dwarf is in view. This will be discussed further in the next section.

\subsection{Characterising the line profiles}

\begin{figure}
\begin{center}
\rotatebox{270}{\includegraphics[height=8.0cm]{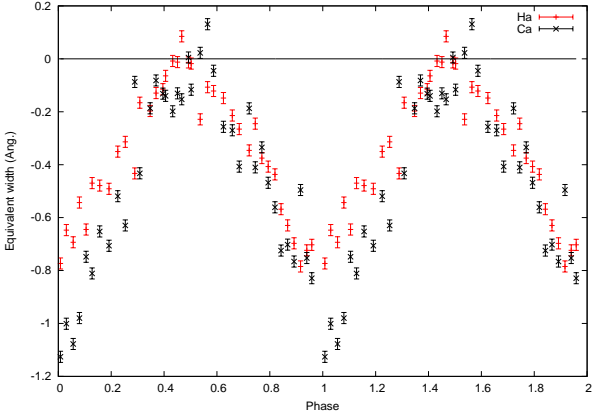}}
\end{center}
\caption{Equivalent width of the H$\alpha$ emission (red '+') and the Ca II 8662\,\AA\, line (black 'x') as a function of phase. The data has been duplicated over two phases.}
\label{fig:ew_plot}
\end{figure}

\begin{figure}
\begin{center}
\rotatebox{270}{\includegraphics[height=8.0cm]{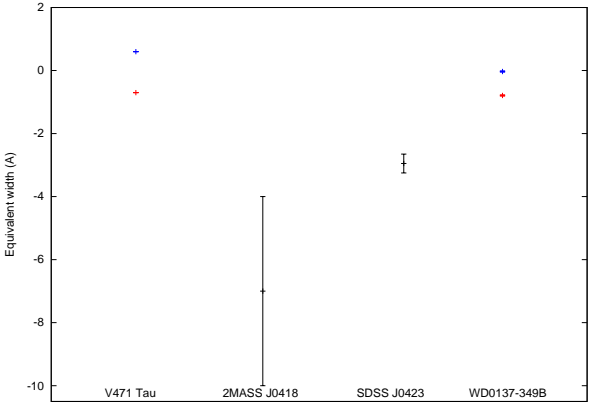}}
\end{center}
\caption{A comparison of our measured equivalent widths to other brown dwarfs that display this emission. V471 Tau has also been included as it has day (red, lower) and night (blue, higher) side equivalent widths with a clear phase dependency similar to WD0137-349B.}
\label{fig:EWcomparison}
\end{figure}

%WD0137-349B has the complex atmosphere of a brown dwarf and is irradiated in a similar way to a hot Jupiter. 

The equivalent widths (EWs) listed in Table~\ref{table:orbprop} demonstrate that there is a clear and dramatic difference between the day-- and night--side of the brown dwarf which fits with the 500\,K temperature difference found by \citet{casewell15}. Figure~\ref{fig:ew_plot} shows the variability in EW of the H$\alpha$ emission line with respect to the phase of the system. EW is effectively a measure of line strength and there is a clear weakening of the line at $\phi = 0.5$: where the dark side of the brown dwarf is facing the observer. The same has been performed for the 8662\,\AA\,Ca II line and is also plotted in Figure~\ref{fig:ew_plot}; the shape of this variation for these two lines matches closely and the Ca II line appears only marginally weaker. It is likely that the phase dependent variability of the brown dwarf indicates significantly different atmospheric properties between the day-- and night--side. This may give rise to extreme variations in weather \citep{lee15}. From \citet{burleigh06} we see that the brown dwarf dominates the continuum at wavelengths longwards of 1.95\,$\mu$m therefore lines detected in the NIR are more largely affected and caution should be exercised when interpreting these EWs.

We have compared our measured EWs to published EWs of two "hyperactive" brown dwarf systems (2MASS J0418 and SDSS J0423) that display H$\alpha$ emission. There are other brown dwarfs with published EWs for this line however they are most often early L dwarfs on or around the hydrogen burning mass limit or the H$\alpha$ emission is sporadic and probably due to flaring. The two included in Figure~\ref{fig:EWcomparison} have sustained H$\alpha$ emission. We also compare to V471 Tau as it is the only other irradiated companion, to our knowledge, whose published EWs display a clear phase dependency. Comparisons to the hyperactive brown dwarfs is valid as the underlying continuum is unaffected by non -- brown dwarf components such as reflection from a companion. Setting the continuum level in V471 Tau was more complex due to line blending and rotational broadening but \citet{bois1991} accounted for this. Thus we can compare our H$\alpha$ values to these literature values. Data is sparse thus any conclusions drawn are preliminary and more observations are required to build a better picture of irradiated atmospheres. That said, Figure~\ref{fig:EWcomparison} suggests that the emission from WD0137-349B is more akin to irradiation induced emission like that seen on V471 Tau than emission seen in "hyper-active" brown dwarfs which is driven by unusually strong magnetic fields.

\begin{figure}
\begin{center}
\rotatebox{270}{\includegraphics[height=8.0cm]{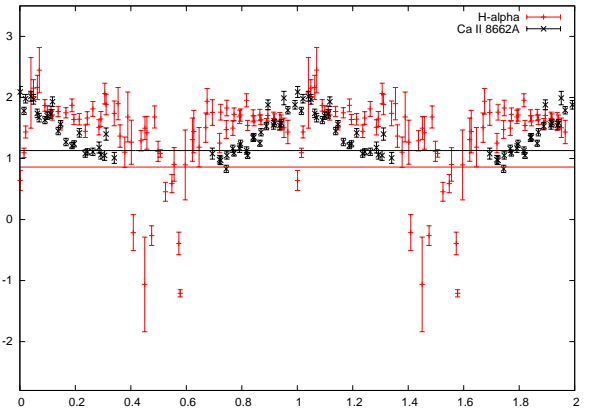}}
\end{center}
\caption{The FWHM of the H$\alpha$ line (red '+') and the Ca II line (black 'x') with respect to phase. The horizontal lines indicate the FWHM calculated from the vsin(i) for each line in the corresponding colour. This has been duplicated over two orbits for clarity.}
\label{fig:fwhm_plot}
\end{figure}

\begin{figure*}
\subfigure[Trailed spectra of the Ca II 8662\,\AA\, emission line.]{\includegraphics[height=8.0cm, width=8.0cm]{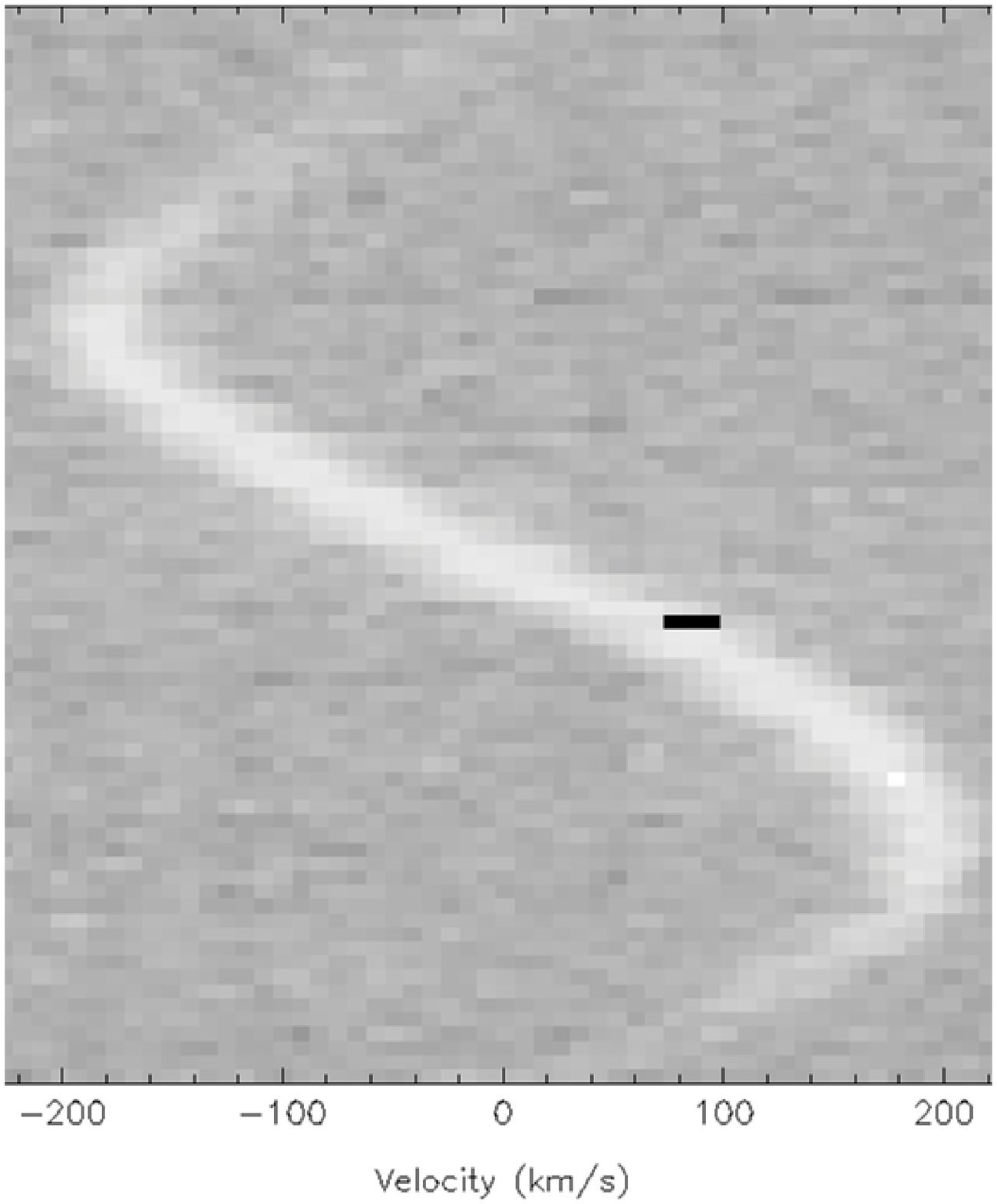}}
\subfigure[Simulated data]{\includegraphics[height=8.0cm, width=8.0cm]{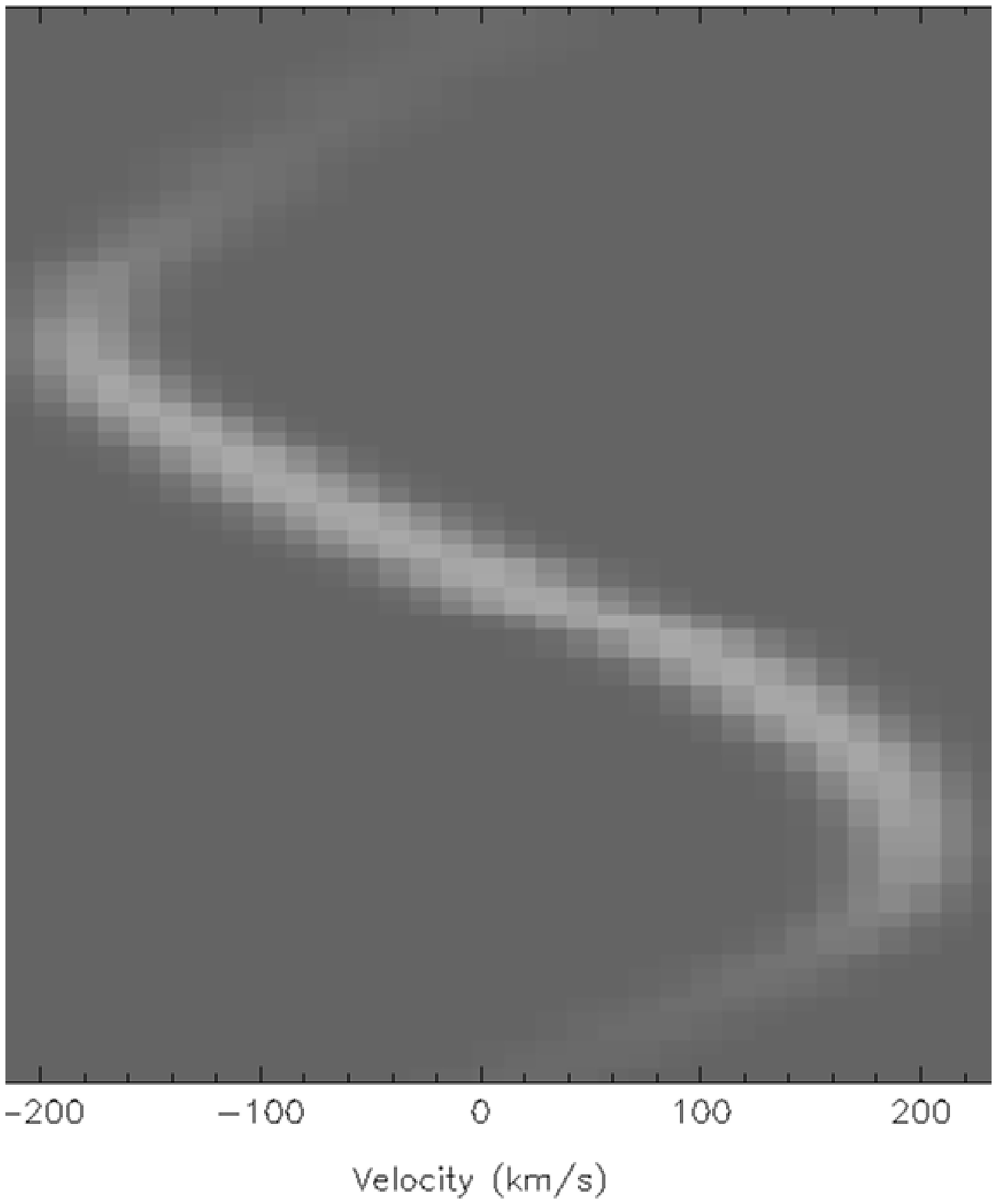}}
\caption{Trail plots of the Ca II data and our simulated Ca II 8662\,\AA\, emission line for comparison.}
\label{fig:radfit}
\end{figure*}

We also observe a phase dependent change in the width of the emission lines as is demonstrated in Figure~\ref{fig:fwhm_plot}. From this we can see that the Ca II emission line is narrower at nearly every point in phase suggesting it may be coming from a region lower in the brown dwarf atmosphere. The signal-to-noise ratio of the Ca II line between $\phi=0.35 - 0.7$ was too low to fit.

To investigate this further we have simulated the Ca II emission lines using a model that takes into account the Roche geometry of the brown dwarf and limb darkening / brightening. We simulate an intrinsic spectrum at each point on the brown dwarf that is equal to Gaussians with the same FWHM as the resolution of the spectra. The emission flux is proportional to the incident flux from the white dwarf and we assume no intrinsic emission or absorption from the brown dwarf i.e. pure emission due to irradiation. The exposure time of 5\,mins is an appreciable fraction of the orbital period, to account for this we have included phase smearing by numerical integration over three phase points.

We set up a grid of semi-amplitude, K, filling factor, f, and linear limb darkening, X, as defined in \citet{maxted98}. We use Levenburg-Marquadt minimisation to optimise $\gamma$ and the ratio of the Ca II intrinsic line strengths plus an overall scaling factor to match the line strengths of the observations. From Figure~\ref{fig:radfit} we can see that these simulations provide a good fit to the data, the best fitting values are listed in Table~\ref{table:masses} along with our calculated mass ratio. We also estimate the v sin(i) of this system to be 39.3\,kms$^{-1}$ which can be translated to FWHM using, FWHM $= \frac{v\,sin(i)}{c} \lambda $ where $\lambda$ is the wavelength of the line. This has been indicated with horizontal lines on Figure~\ref{fig:fwhm_plot} for H$\alpha$ and Ca II. \\

\begin{table}		%Masses
\begin{center}
\begin{tabular}{l | c | c | r}
\toprule
\multicolumn{4}{|c|}{Properties of the system} \\
\midrule
P (days) & &									& $ 0.079429939(1)$		\\
T$_0$ & &										& $ 2454178.6761\pm0.0003 $ \\
K$_{\rm c}$ (km\,s$^{-1}$)& & 	& $ 210 \pm 5 $ \\
Mass ratio		& &							& $ 0.135 \pm 0.004 $		\\
\bottomrule
\end{tabular}
\end{center}
\caption{The values obtained from fitting equation~(\ref{eq:rv}) to the measured radial velocities of the H$\alpha$ absorption and emission lines. The error on the final digit of the period is shown in parentheses. We list the corrected semi-amplitude for the brown dwarf used to correct for the light from the emission being offset from the centre of mass of the brown dwarf; we used this value to calculate the masses of the white dwarf and brown dwarf.}
\label{table:masses}
\end{table}

\noindent The work presented in this paper demonstrates the first brown dwarf atmosphere to exhibit a range of emission lines as a direct result of intense UV irradiation from the 16\,500\,K white dwarf primary. Ability to reproduce the line strength variation that is seen in WD0137-349 will be important in any future models of irradiated atmospheres.

\subsection{Comparison to hot Jupiter models}

In an attempt to explore the role irradiation and opacity plays on heat redistribution \citet{perna12} model the atmosphere of hot Jupiters using 3D atmospheric circulation models. The models are cloud-free and this results in an atmosphere opacity that is too low. They do find however, that heat redistribution breaks down at irradiation temperatures greater than 2200 - 2400\,K and attribute this to the advective timescale becoming much longer than the radiative timescale. \citet{showman13} use a circulation model to compare the atmospheres of hot Jupiters with and without intense irradiation. They found that strong insolation (stellar radiation that reaches the planets surface) produces circulation dominated by high-altitude air flow from the day--side to the night--side and suggest a return flow at deeper levels. It is clear irradiation of gas giant planets introduces additional dynamical effects within their atmospheres. Younger, lower gravity (log g $\approx$ 3) brown dwarfs share many properties with these gas giant exoplanets \citep{faherty13}. 

No 3D simulations have been created for higher gravity irradiated brown dwarfs like WD0137-349B but we may expect comparable atmospheric effects. A recent study by \citet{hern16} looks at the interacting binary system of a brown dwarf being irradiated by a 13\,200\,K white dwarf. They model the atmosphere using a simple geometric reprocessing model and find poor heat redistribution from the day-- to the night--side. With a slightly hotter white dwarf such as WD0137-349 (T$_{\rm eff}$ = 16\,500\,K) one would expect this effect to be similar and the strength variation over the orbit of all our observed emission lines does support this. Although the longer period of WD0137-349B is likely to make heat redistribution more effective. Full analysis of phase curves would need to be done to be sure of this effect.

\section{Conclusion}

We have used 78 XSHOOTER observations of WD0137-349AB to refine the ephemeris of the system using the emission and absorption lines of H$\alpha$. We show that this system is detached as we see no indication of mass transfer in the trailed spectra and no lines other than balmer lines coming from the white dwarf. We have detected new emission lines of He~I, Na~I, Mg~I, Si~I, K~I, Ca~I \& II, Ti~I, and Fe~I \& II originating from the brown dwarf atmosphere as a direct result of irradiation.

WD0137-349 is unique as it is the first brown dwarf to exhibit irradiation induced metal emission lines. One would expect that shorter orbits with hotter white dwarfs, thus more intense irradiation, would lead to stronger Balmer line emission and metal emission like we see in the atmosphere of WD0137-349B. We have discussed comparable systems and find that despite their similarities they do not all manifest the same observable properties.

To establish what to expect from an irradiated brown dwarf atmosphere we use a {\sc drift-phoenix} model and manipulate the temperature of the upper atmosphere to give an outward temperature increase. This provides a simple preliminary test of the expected chemical composition of a brown dwarf atmosphere at different gas temperatures or atmospheric heights. The elements we have detected and their ionisation states loosely match the predictions of this model which supports the possibility of a brown dwarf chromosphere being present.

We measured the equivalent widths of the emission lines in WD0137-349B and found the line strengths to weaken from the day to the night--side of the brown dwarf. This indicates that the irradiation is likely driving dramatic temperature and weather variations between the two sides of this tidally locked brown dwarf. We found the H$\alpha$ emission line to have a strong phase dependency and compared our H$\alpha$ day-- and night--side equivalent width to that of similar systems. We find that WD0137-349B has equivalent widths more akin to the irradiation induced H$\alpha$ emission of low mass stellar companion atmospheres. 

Any future models of irradiated atmospheres will need to reproduce the line strength variation that is seen here.

\section*{Acknowledgements}

We would like to thank our helpful reviewer for their comments and suggestions. We also thank Tom Marsh for the use of \textsc{molly}. E. S. Longstaff acknowledges the support of STFC studentship. S. L. Casewell acknowledges support from the University of Leicester College of Science and Engineering. Ch. Helling highlights the financial support of the European community under the FP7 ERC starting grant 257431. This work was supported by the Science and Technology Facilities Council [ST/M001040/1]. This work is based on observations made with ESO telescopes at La Silla Paranal Observatory under programme ID 093.C-0211(A). 

%%%%%%%%%%%% REFERENCES %%%%%%%%%%%%

\bibliographystyle{mnras}
\bibliography{references} % if your bibtex file is called references.bib 

% Don't change these lines
\bsp	% typesetting comment
\label{lastpage}
\end{document}